\documentclass[article,a4paper,twocolumn,accepted=2025-01-16,]{quantumarticle}
\pdfoutput=1
\AddToHook{begindocument/before}{\usepackage{hyperref}}
\usepackage{amsmath}
\usepackage{graphicx}
\usepackage{physics}
\usepackage{amssymb}
\usepackage{amsthm}
\usepackage[numbers,sort&compress]{natbib}

\newtheorem{theorem}{Theorem}

\begin{document}

\title{Distance-preserving stabilizer measurements in hypergraph product codes}

\author{Argyris Giannisis Manes}
\email{agiannisismanes@uchicago.edu}
\author{Jahan Claes}
\email{jahan.claes@yale.edu}
\affiliation{Departments of Physics and Applied Physics, Yale University, New Haven, CT 06520, USA\\Yale Quantum Institute, Yale University, New Haven, Connecticut 06511, USA}

\begin{abstract}
Unlike the surface code, quantum low-density parity-check (QLDPC) codes can have a finite encoding rate, offering potentially lower overhead. However, finite-rate QLDPC codes have nonlocal stabilizers, making it difficult to design stabilizer measurement circuits that are low-depth and preserve the effective distance. Here, we demonstrate that a popular {construction} of finite-rate QLDPC codes, the hypergraph product, has the convenient property of distance-robustness: any stabilizer measurement circuit preserves the effective distance. In particular, we prove a previously proposed depth-optimal circuit is also optimal in terms of effective distance.
\end{abstract}

\maketitle

\section{Introduction}

Building a large-scale quantum computer will likely require encoding information in quantum error-correcting codes. Quantum low-density parity-check (QLDPC) codes are a broad family of quantum {stabilizer} codes in which errors are detected by measuring Pauli stabilizers {of weight $O(1)$}, and each qubit is included in {$O(1)$} stabilizers~\cite{breuckmann2021quantum}. A quantum code on $n$ physical qubits encoding $k$ logical qubits that can detect up to $(d-1)$ errors is denoted an $[[n,k,d]]$ code. The most widely studied QLDPC code is the surface code~\cite{kitaev1997quantum,bravyi1998quantum}, in which qubits are arranged in 2D {Euclidean} space so the stabilizers are geometrically local, and whose $[[n,k,d]]$ parameters satisfy $kd^2=O(n)$. Indeed, it has been proven that any local QLDPC code~{in 2D Euclidean space} necessarily has $kd^2=O(n)$~\cite{bravyi2009no,bravyi2010tradeoffs}, and improving on these parameters requires some amount of non-local connectivity~\cite{baspin2022quantifying,baspin2023improved}~{or non-Euclidean geometry~\cite{breuckmann2016constructions}}. Thus, {to achieve} a finite-rate code (in which $k=O(n)$ and $d>O(1)$), we must go beyond local 2D codes. \footnote{{Note if we do not have $d>O(1)$, the fraction of errors that are correctable will vanish as $n\rightarrow\infty$}.}

There have been several recent constructions of {non-local} QLDPC codes with increasingly better parameters~\cite{freedman2002z2,breuckmann2016constructions,guth2014quantum,londe2017golden,tillich2013quantum,hastings2020fiber,breuckmann2021balanced,panteleev2022asymptotically,leverrier2022quantum,lin2022good}, culminating in the construction of ``good" QLDPC codes with $k,d=\Theta(n)$~\cite{panteleev2022asymptotically}{\cite[see also][]{leverrier2022quantum,lin2022good}}. However, there is still considerable interest in hypergraph product (HGP) codes~\cite{tillich2013quantum}, an earlier {construction achieving} $k=\Theta(n)$, $d=\Theta(\sqrt n)$. While HGP codes have worse $d$ scaling than ``good" QLDPC codes, the scaling for HGP codes is sufficient to prove a threshold exists~\cite{kovalev2013fault,gottesman2013fault} and that large-scale quantum error correction requires only a constant overhead factor~\cite{gottesman2013fault}. In addition, the HGP codes' relatively tractable construction has led to further development of decoders~\cite{leverrier2015quantum,fawzi2018efficient,grospellier2018numerical,roffe2020decoding,grospellier2021combining,quintavalle2022reshape,crest2023layered}, logical gates~\cite{krishna2021fault,quintavalle2022partitioning,cohen2022low}, and low-depth stabilizer measurement circuits~\cite{tremblay2022constant,delfosse2021bounds}, and it has been recently demonstrated that HGP codes can be designed to be compatible with modular architectures~\cite{strikis2023quantum} and reconfigurable atom arrrays~\cite{xu2023constant}.

For any QLDPC code, an immediate question is: how do we measure the stabilizers? One typically wants to measure stabilizers using ancilla qubits and low-depth quantum circuits, but these circuits must be optimized to simultaneously measure overlapping stabilizers~\cite{dennis2002topological,fowler2012surface,landahl2011fault,beverland2021cost}. Ref.~\cite{tremblay2022constant} introduced low-depth circuits for general CSS-type QLDPC codes, as well as even lower-depth circuits that measured all stabilizers in parallel for particular HGP codes. Besides the depth of measurement circuits, one should also optimize how measurement circuits propagate errors. A single error on an ancilla qubit during a circuit measuring a weight-$\omega$ stabilizer can propagate to a weight $\lfloor \omega /2\rfloor$ error on the data qubits, which may reduce the effective distance of the code from $d$ to $\lceil d/\lfloor \omega/2\rfloor \rceil$ in the worst case~\cite{tomita2014low}. The original surface code has the convenient property that no measurement circuit reduces the effective distance~\cite{dennis2002topological}, but the rotated surface code requires a carefully chosen measurement schedule to avoid reducing the distance to $\lceil d/2\rceil$~\cite{tomita2014low}. For a general QLDPC code with many logical operators and no local structure, there may not exist a circuit that preserves the distance. Indeed, a recent work~\cite{bravyi2023high} has proposed stabilizer measurement circuits for a finite-rate QLDPC code family closely related to HGP codes known as quasicyclic codes~\cite{kovalev2013quantum}. In that work, it was found that randomly chosen stabilizer measurement circuits sharply reduced the code distance, and even optimized circuits could not fully preserve the distance~\cite{bravyi2023high}.

In this paper, we prove that HGP codes have the same distance robustness as the original surface code: any valid stabilizer-measurement circuit does not reduce the effective distance of the code. As HGP codes are a generalization of the surface code, our result is a generalization of the robustness of the surface code. Our result shows that the ``product coloration circuit" that is most directly compatible with reconfigurable atom arrays~\cite{xu2023constant} can be used without reducing the distance. In addition, our result establishes that the ``cardinal" measurement circuit for HGP codes introduced in~\cite{tremblay2022constant} is in some sense optimal: the circuit has the lowest possible depth and does not reduce the effective code distance, satisfying the two typical desiderata for measurement circuits~\cite{landahl2011fault,tomita2014low}.

{A natural question raised by our proof is: Can it be applied to the lifted product~\cite{panteleev2021quantum,panteleev2022asymptotically}, a generalization of the HGP with linear distance scaling? Unfortunately, for general lifted product codes, no such proof is possible. Lifted product codes contain all generalized bicycle codes~\cite{panteleev2021quantum}, and generalized bicycle codes contain both a subset of rotated toric codes which are not distance robust~\cite[][example 2]{kovalev2013quantum} as well as the above-mentioned quasicyclic codes~\cite{bravyi2023high}. We note that it is still an interesting question whether some subset of lifted products can be found with both linear distance scaling and distance-robustness.}

\section{Preliminaries: HGP Construction}

\begin{figure}
    \centering
    \includegraphics[width=\columnwidth]{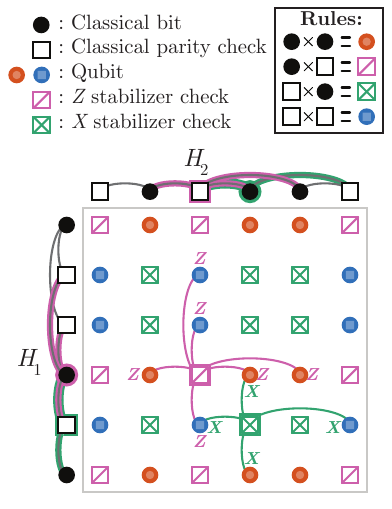}
    \caption{Lattice illustration of an HGP code as the Cartesian product of two Tanner graphs of classical linear codes $H_1,H_2$. The black {circles} and squares correspond to the bits and checks of the classical codes. The product of a bit with a bit (check with a check) is a bit-type (check-type) qubit, illustrated in orange (blue). The product of a bit with a check (check with a bit) is a $Z(X)$-stabilizer. The stabilizers are supported on {connected} qubits in both directions; {the connections} are determined from the classical parity check matrices. Two example stabilizers are illustrated.}
    \label{fig:HGPstab}
\end{figure}

The HGP construction~\cite{tillich2013quantum} takes two classical linear codes, described {by} parity check matrices $H_1$ and $H_2$ with $n_i$ {bits} and $r_i$ parity checks, and defines a quantum CSS code~\cite{calderbank1996good,steane1996multiple}. {Note that the rows of $H_i$ may be linearly dependent, i.e. we allow redundant checks.} We illustrate the construction in Fig.~\ref{fig:HGPstab}. Each classical code {is} described by a Tanner graph, where bits are represented by circles and parity checks are represented by squares {connected to the circles they act on}. To form the HGP code, we take the Cartesian product of the two classical Tanner graphs. The qubits of the quantum code come in two types, given by the product of the bits and bits or checks and checks (hereafter, bit- and check-type qubits). The $X(Z)$-stabilizers of the code are given by the product of checks and bits (bits and checks), and act on the qubits they are connected to in the Cartesian product graph. The $X$- and $Z$-stabilizers are then guaranteed to commute. Note that every stabilizer~{in Fig.~\ref{fig:HGPstab}} acts on the qubits in the same row or column as that stabilizer. $X(Z)$ stabilizers act on bit(check)-type qubits above and below them, and check(bit)-type qubits to their left and right.

Algebraically, this construction corresponds to check matrices {$H_X(H_Z)$} whose rows describe the {$X(Z)$}-stabilizers
\begin{equation}
\begin{split}
    H_X &= \left(\begin{matrix}H_1 \otimes I_{n_2} &, & I_{r_1} \otimes H_2^T \end{matrix}\right) \\
    H_Z &= \left(\begin{matrix} I_{n_1} \otimes H_2 & , & H_1^T \otimes I_{r_2} \end{matrix}\right)
\end{split}
\label{eq:matrices_standard}
\end{equation}
{where the comma separates concatenated matrices.} The $n_1n_2$ columns {before the comma} correspond to the bit-type qubits, and the remaining $r_1r_2$ columns {after the comma} correspond to the check-type qubits.

The parameters $[[n,k,d]]$ of the HGP code can be expressed in terms of the classical code parameters. {Let} the classical code $H_i$ {have} parameters $[n_i,k_i,d_i]${; }let the classical transposed code $H_i^T${,} formed by swapping the role of checks and bits in the Tanner graph{, have} parameters $[r_i,k_i^T,d_i^T]$. Then the HGP {code} has
\begin{align}
n&=n_1n_2+r_1r_2\\
k&=k_1k_2+k_1^Tk_2^T\\
d&=\left\{\begin{array}{ll}
\min(d_1,d_2) &,\  d_{1}^T = \infty\text{ or }d_2^T = \infty\\
\min(d_1^T,d_2^T) &,\  d_1=\infty\text{ or }{d_2}=\infty\\
\min(d_1,d_2,d_1^T,d_2^T) & ,\ \text{else}.
\end{array}\right.
\end{align}
The formula for $n$ is obvious, while $k$ can be straightforwardly derived by computing the ranks of $H_X$ and $H_Z$. We prove the formula for $d$ while proving Thm.~\ref{thm:DistanceRobustness} {below}. Choosing {classical LDPC} codes with $k_i,d_i\propto n_i$ gives sparse HGP codes with $k\propto n$ and $d\propto\sqrt{n}$~\cite{tillich2013quantum}. 

\section{HGP Logical Operators}

{To understand} the distance of the HGP code, we explicitly construct the $k$ logical operators. These logical operators are formed by taking the product of classical codewords of {$H_i$ or $H_i^{T}$} with {certain weight-one vectors}. 

\begin{figure}
    \centering
    \includegraphics[width=\columnwidth]{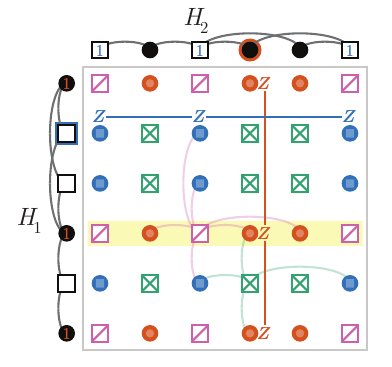}
    \caption{Examples of elementary logical $Z$ operators for an HGP code, as given in Eqs~\ref{eq:firstZLogicals} and~\ref{eq:secondZLogicals}. The logical operators are obtained by combining a codeword of one of the classical codes (in this case, $111$) with a unit vector that lies outside the image of the other code's transpose check matrix.}
    \label{fig:logical}
\end{figure}

Following~\cite{tillich2013quantum,quintavalle2022reshape}, we define two classes of logical $Z$ operators, illustrated in Fig.~\ref{fig:logical}. The first is generated by a basis $\{x_i\}$ of classical codewords of $H_1$ and a set $\{y_j\}$ of unit vectors whose linear combinations are outside the image of $H_2^T$. Clearly, $|\{x_i\}|=k_1$. Since $\dim\left[\Im(H_2^T)\right]=(n_2-k_2)$, we can always find $k_2$ unit vectors whose span never falls inside $\Im(H_2^T)$, {so that} $|\{y_j\}|=k_2$. {The first} $k_1k_2$ logical $Z$ operators {are then} given by
\begin{equation}
    \left(\begin{matrix}x_i\otimes y_j\\0\end{matrix}\right).\label{eq:firstZLogicals}
\end{equation}
These operators are oriented vertically {in Fig.~\ref{fig:logical}} and act only on bit-type qubits. They have weight $\geq d_1$, with at least {$k_1$} having weight $d_1$ (provided $k_2\neq 0$). Pictorially, it's clear they commute with $X$-stabilizers: each $X$-stabilizer intersects the $Z$ operator either zero times, or the same number of times its corresponding classical check in $H_1$ intersects the classical codeword $x_i$. Because $x_i$ is a codeword of $H_1$, this intersection must be even.

The second class is generated by a basis $\{b_m\}$ of classical codewords of $H_2^T$ and a set $\{a_\ell\}$ of unit vectors whose linear combinations are outside $\text{Im}(H_1)$. We {similarly} have $|\{b_m\}|=k_2^T$, $|\{a_\ell\}|=k_1^T$. Then the second $k_1^Tk_2^T$ {logical $Z$ operators are given by}
\begin{equation}
    \left(\begin{matrix}0\\ a_\ell \otimes b_m\end{matrix}\right).\label{eq:secondZLogicals}
\end{equation}
which are oriented horizontally and act only on check-type qubits. They have weight $\geq d_2^T$, with at least {$k_1^T$} having weight $d_2^T$ (provided $k_2^T\neq 0$). These operators commute with $X$-stabilizers for similar reasons. We have thus enumerated $k=(k_1k_2+k_1^Tk_2^T)$ logical $Z$-operators.

{To ensure that these $Z$-operators are independent, we need to establish that no combination of $Z$-operators can be written as a combination of $Z$ stabilizers. WLOG, we consider a combination of $Z$ operators which acts on at least one row of bit-type qubits  (e.g., highlighted row of Fig.~\ref{fig:logical}). We can restrict to this row and view it as a copy of the classical code $H_2$. The combination of $Z$-operators restricted to this row is made up of a combination of $\{y_j\}$. Any $Z$ stabilizer restricted to this row is an element of $\text{Im}(H_2^T)$. Thus, since no combination of $\{y_j\}$ is in $\text{Im}(H_2^T)$, no combination of $Z$ stabilizers can equal the logical operator.}

{We summarize this discussion in the following theorem and formal proof:}

\begin{theorem}[\cite{tillich2013quantum,quintavalle2022reshape}]
The set of operators given by Eqs.~\ref{eq:firstZLogicals} and~\ref{eq:secondZLogicals} are a complete set of independent logical $Z$ operators.
\end{theorem}

\begin{proof}
    {We need to show that the set of operators given by Eqs.~\ref{eq:firstZLogicals} and~\ref{eq:secondZLogicals} are a basis for $\text{ker}(H_X)/\text{Im}(H_Z^T)$. We note that this space is $k$-dimensional, so if the $k$ vectors enumerated by Eqs.~\ref{eq:firstZLogicals} and~\ref{eq:secondZLogicals} are in $\text{ker}(H_X)$ and are linearly independent in $\text{ker}(H_X)/\text{Im}(H_Z^T)$, they are a basis. We therefore consider an arbitrary nontrivial linear combination of such operators:} 
    \begin{equation}
        z = \left(\begin{matrix}\sum_{ij} \lambda_{ij} x_i \otimes y_j \\
        \sum_{\ell m} \kappa_{\ell m} a_\ell \otimes b_m\end{matrix}\right){,}
    \end{equation}
    {where $\lambda_{ij}$ and $\kappa_{\ell m}$ are arbitrary binary values}.

    First, we show $z\in\text{ker}(H_X)$. We have (see Eq.~\ref{eq:matrices_standard} for $H_X$)
    \begin{equation}
    \begin{split}
        H_X z &= {\sum_{ij} \lambda_{ij} \left(H_1x_i\right) \otimes y_j +\sum_{\ell m} \kappa_{\ell m} a_\ell \otimes \left(H_2^T b_m\right)}\\&=0
    \end{split}
    \end{equation}
    where we have used $H_1x_i=0$ ($H_2^Tb_m=0$), since $x_i$ ($b_m$) are codewords of $H_1$ ($H_2^T$).

    {Second, we show that $z\notin\text{Im}(H_Z^T)$, so that $z$ is nonzero in $\text{ker}(H_X)/\text{Im}(H_Z^T)$. We establish this by contradiction. If $z\in\text{Im}(H_Z^T)$}, there would exist some vector $w$ such that $z = H_Z^T w$, or
    \begin{equation}
        z=\left(\begin{matrix}\sum_{ij} \lambda_{ij} x_i \otimes y_j \\
        \sum_{\ell m} \kappa_{\ell m} a_\ell \otimes b_m\end{matrix}\right) = \left(\begin{matrix}(I_{n_1}\otimes H_2^T)w\\(H_1\otimes I_{r_2})w\end{matrix}\right)\label{eq:ContradictionAssumption}
    \end{equation}
    We can define new codewords $x_j':=\sum_i \lambda_{ij}x_i$ and $b_\ell':=\sum_m \kappa_{\ell m}b_m$. We note that if some $\lambda_{ij}\neq 0$ then the corresponding $x_j'\neq 0$, since the $x_i$ are linearly independent. Similarly, if some $\kappa_{\ell m}\neq 0$ then the corresponding $b_\ell'\neq 0$. Without loss of generality, let's assume $\lambda_{ij}\neq 0$ and $x_j'\neq 0$; the case where all $\lambda_{ij}=0$ and some $\kappa_{\ell m}\neq 0$ can be treated similarly. Expand $w$ as $w=\sum_k w_k\otimes \tilde{w}_k$ {(such a decomposition is always possible by choosing $\{w_k\}$ to be a basis of the first tensor product factor)}. Then the top {row} of Eq.~\ref{eq:ContradictionAssumption} becomes
    \begin{equation}
        \sum_j x_j'\otimes y_j = \sum_k w_k\otimes \left(H_2^T \tilde{w}_k\right).
    \end{equation}

    {As there exists an $x_j'\neq 0$, it is possible to pick an $n_1$-dimensional unit vector $e_t$ such that $e_{t}\cdot x_j'\neq 0$. If we apply the linear operator $e_{t}^T\otimes I_{n_2}$ (an $n_2\times n_1n_2$ dimensional matrix) to both sides, we have}
    \begin{equation}
        \sum_j (e_{t}\cdot x_j')y_j = \sum_k (e_{t}\cdot w_k)H_2^T \tilde{w}_{k}
    \end{equation}
    which {says a nonzero combination of $\{y_j\}$ is in $\text{Im}(H_2^T)$, contradicting our choice for the $\{y_j\}$}. Thus $z\notin \text{Im}(H_Z^T)$.
\end{proof}

The logical $X$ operators can be found similarly, by swapping the roles of $H_1$ and $H_2$.

\section{Effective Code Distance}

\begin{figure}[h!]
    \centering
    \includegraphics[width=\columnwidth]{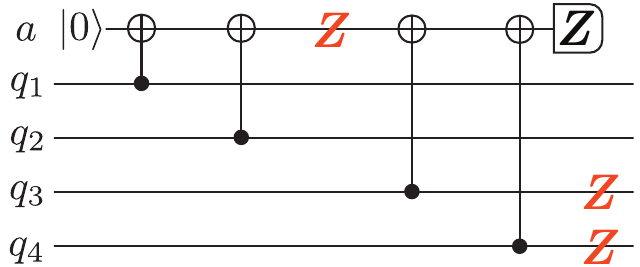}
    \caption{An example circuit for measuring a $Z$ stabilizer supported on code qubits $q_1,\dots,q_4$. A single $Z$ error on the ancilla qubit halfway through the circuit propagates to $Z$ errors on code qubits $q_3$ and $q_4$, potentially {reducing the number of physical errors that can be corrected}.}
    \label{fig:measurement}
\end{figure}

When measuring a stabilizer using a single ancilla qubit and a set of two-qubit entangling gates, errors can propagate from the ancilla qubit to multiple code qubits. For instance, consider the circuit measuring a $Z$ stabilizer of weight $4$ in Fig. \ref{fig:measurement}.  In this case, if a $Z$ error occurs on the ancilla qubit during the measurement, it can propagate to two physical qubits. Such an error is often called a ``hook error" in the literature~\cite{dennis2002topological}

In general, when measuring a stabilizer of weight $\omega$, a single ancilla error can propagate to any subset of code qubits in the support of the stabilizer (the specific subsets will depend on the order we apply the two-qubit gates). Because we can always multiply an error by a stabilizer to reduce its weight, a single error will propagate to up to $\lfloor \frac{\omega}{2} \rfloor $ code qubits. This error propagation may reduce {the number of physical errors required for an undetectable logical error by a factor of up to $\lfloor \frac{\omega}{2} \rfloor $}~\cite{landahl2011fault,tomita2014low,beverland2021cost}.

{For a given code and measurement circuit, we define the{ \bf effective distance} to be the minimum number of physical errors required for an undetectable logical error. We note that only ancilla errors propagating to data qubits can reduce the effective distance, as data qubit errors never spread to neighboring data qubits.}.

{I}n the particular case of HGP codes, error propagation {never reduces} the distance of the code. Specifically, if we assume {we measure each row of $H_Z(H_X)$ using a circuit as in Fig.~\ref{fig:measurement}} and allow a $Z(X)$ error on the ancilla qubit to introduce a $Z(X)$ error on any subset of the qubits in the support of that stabilizer, we will show that at least $d$ {physical} errors are {still} needed {for an undetectable} logical error\footnote{{Note that a $Y$ error on an ancilla qubit propagates as either an $X$ or $Z$ error during measurement, so we do not need to separately consider $Y$ errors.}}. This is summarized in the following theorem. 

\begin{theorem}
    Let $d$ be the distance of an HGP code. The effective distance of the code using any stabilizer measurement {order for the rows of $H_Z$ and $H_X$} is also $d$.\label{thm:DistanceRobustness}
\end{theorem}

{Before formally proving this statement, we provide an intuitive explanation. Any combination $z$ of logical operators from Eqs.~\ref{eq:firstZLogicals} and~\ref{eq:secondZLogicals} will have at least one column (row) with at least $d_1$ ($d_2^T$) $Z$ operators. WLOG, assume it is a column, as in the vertical logical operator in Fig.~\ref{fig:logical}. $z$ then has support on at least $d_1$ rows. Restricted to any one of these rows, $z$ is a combination of vectors $\{y_i\}$. Therefore, any combination of $z$ with stabilizers still has support on this row, since any stabilizer restricted to this row is in $\text{Im}(H_2^T)$ and a combination of $\{y_i\}$ can never be in $\text{Im}(H_2^T)$. Thus, even if we multiply $z$ by $Z$ stabilizers, we cannot have a logical operator with support in fewer than $d_1$ rows. If we multiply $z$ by a partial $Z$ stabilizer induced by an ancilla error, at worst this stabilizer can cancel the support of $z$ on one row, since each stabilizer only acts on a single row of bit-type qubits. Thus, an ancilla error can only reduce the weight of the logical by $1$, just like a direct data qubit error.}

We now turn to our formal proof.

\begin{proof}[Proof (See also~{\cite[Prop. 2]{quintavalle2022reshape}}).]
    An elementary $Z$ stabilizer {(row of $H_Z$)} is given by $s_{t\tilde t}:=H_Z^T(e_t\otimes \tilde e_{\tilde t})$, where $e_{t}\otimes \tilde e_{\tilde t}$ is a unit vector, so
    \begin{equation}
        s_{t\tilde t} = \left(\begin{matrix}
        e_t \otimes H_2^T \tilde{e}_{\tilde t} \\ H_1 e_t \otimes \tilde{e}_{\tilde t}
        \end{matrix}\right).
    \end{equation}
    An ancilla error may propagate $Z$ errors to a subset of the qubits in the support of the stabilizer. However, we see that a stabilizer $s_{t\tilde t}$ is supported on qubits $e_t\otimes (\cdot)$ and $(\cdot)\otimes \tilde{e}_{\tilde t}$. Pictorally, this says that a $Z$ stabilizer $s_{t\tilde t}$ only act on bit-type qubits in the row $e_t$ and check-type qubits in the column $\tilde{e}_{\tilde t}$, as illustrated in Fig.~\ref{fig:HGPstab}. If every logical $Z$ operator has support in at least $d_1$ distinct rows $e_t$ or $d_2^T$ distinct columns $\tilde{e}_{\tilde t}$, we still require $d$ ancilla errors to cover a logical operator.
    
    Consider an arbitrary logical $Z$ operator consisting of a product of logical $Z$ operators and $Z$ stabilizers
    \begin{equation}
    z= \left( \begin{matrix}
                \sum_j x_j' \otimes y_j +\sum_k w_k\otimes H_2^T\tilde w_k\\ \sum_\ell a_\ell\otimes b_\ell'+\sum_k H_1w_k\otimes\tilde{w}_k
                \end{matrix} \right) := \begin{pmatrix}
                    z_1 \\ z_2
                \end{pmatrix}.\label{eq:arbitraryZ}
    \end{equation}
    where we have again made use of the definitions of $x_j'$ and $b_\ell'$ introduced above, and again expanded $w$ as $\sum_k w_k\otimes\tilde{w}_k$. Without loss of generality, assume some $x_j'\neq 0$ (the case where all $x_j'=0$ and some $b_\ell'\neq 0$ proceeds similarly). 
    
    Since $x_j'\neq 0$, {it must have hamming weight} $\geq d_1$ and there are therefore at least $d_1$ distinct unit vectors $e_t$ such that $e_t\cdot x_j'=1$. We again {apply $e_t^T\otimes I_{n_2}$ to the top row of Eq.~\ref{eq:arbitraryZ}}:
    \begin{equation}
        (e_t^T\otimes I_{n_2}) z_1 = \sum_j (e_t\cdot x_j')y_j+\sum_k (e_t\cdot w_k)H_2^T \tilde{w}_k.
    \end{equation}
    This {equation} cannot be zero, because at least one of the $(e_{t}\cdot x'_j)\neq 0$, and no nonzero linear combination of $y_j$ is in the image of $H_2^T$. We then see even when allowing for multiplication by stabilizers, $z_1$ has nonzero support in each row corresponding to $e_{t}$. Thus, we still require $d_1$ physical errors to cover the logical operator $z$.

    The proof for $X$ logical operators is identical up to exchanging the roles of $H_1$ and $H_2$.
\end{proof}

\begin{acknowledgments}
{\it Note added---} After preparing the initial version of our manuscript, we became aware that the fact that logical operators have support in at least $d$ rows/columns has also been independently shown in~\cite[Prop 2]{quintavalle2022reshape} in a different context. However, our overall proof is distinct from theirs, and our application to distance robustness is novel.

{\it Acknowledgments---} We are grateful to Shruti Puri and Shilin Huang for helpful discussions. This material is based on work supported by the National Science Foundation (NSF) under Award No. 2137740. Any opinions, findings, and conclusions or recommendations expressed in this publication are those of the authors and do not necessarily reflect the views of NSF.
\end{acknowledgments}

\bibliographystyle{unsrtnat}
\bibliography{thebibliography.bib}

\end{document}